\documentclass[superscriptaddress,nofootinbib,preprint]{revtex4}
\usepackage{graphicx}
\usepackage{amssymb,amsmath,amsfonts}
\usepackage[hypertex]{hyperref}

\def\lozenge{\boxit{\hbox to 1.5pt{\vrule height 1pt width 0pt \hfill}}}
\def\ie{{\it i.e.}}
\def\eg{{\it e.g.}}

\def\mpl{\ifmmode M_{pl}\else $M_{pl}$\fi}
\def\mpl{\ifmmode \overline M_{Pl}\else $\bar M_{Pl}$\fi}
\def\to{\rightarrow}

\newcommand{\lsim}{\lesssim}
\newcommand{\gsim}{\gtrsim}

\newcommand{\sn}{S^n}
\newcommand{\sii}{S^2}
\newcommand{\vphi}{\varphi}
\newcommand{\dalam}{\raise-1mm\hbox{\large$\Box$}}

\newcommand{\beq}{\begin{equation}}
\newcommand{\eeq}{\end{equation}}

\begin{document}

\pagestyle{plain}

\hfill$\vcenter{ \hbox{\bf BNL-HET-07/4}
\hbox{\bf SLAC-PUB-12327}} $

\vskip 1cm
\title{Signatures of Spherical Compactification at the LHC}

\author{Hooman Davoudiasl}

\email{hooman@bnl.gov}
\affiliation{Department of Physics,
Brookhaven National Laboratory, Upton, NY 11973-5000, USA}
\author{{Thomas G. Rizzo}\footnote{Work supported in part
by the Department of Energy, Contract DE-AC02-76SF00515.}}
\email{rizzo@slac.stanford.edu}
\affiliation{Stanford Linear Accelerator Center, 2575 Sand Hill Rd.,
Menlo Park, CA  94025, USA}

\begin{abstract}

TeV-scale extra dimensions may play an important role in 
electroweak or supersymmetry breaking.  We examine the phenomenology 
of such dimensions, compactified on a sphere
$S^n$, $n \geq 2$, and show that they possess distinct features and
signatures.  For example, unlike flat toroidal manifolds,
spheres do not trivially allow fermion massless modes.
Acceptable phenomenology then generically leads to ``non-universal" extra dimensions with
``pole-localized" 4-$d$ fermions; the bosonic fields can be in the bulk.
Due to spherical symmetry, some
Kaluza-Klein (KK) modes of bulk gauge fields are either stable or
extremely long-lived, depending on the graviton KK spectrum.
Using precision electroweak data, we constrain the lightest gauge
field KK modes to lie above $\simeq 4$~TeV.  We show that some of these KK resonances
are within the reach of the LHC in several different production channels. 
The models we study can be uniquely identified by their collider signatures.

\end{abstract}
\maketitle


\section{Introduction}

A great deal of attention has been devoted to the theoretical
development of models with extra dimensions over the past several
years.  Motivated by a desire to explain the gauge hierarchy problem
in the Standard Model (SM), various scenarios with one or more extra
dimensions have been proposed.  These models generally give rise to new
phenomena not far above the weak scale.  Nearly all cases that have been
studied are endowed with extra dimensions that are: {\it (1)} large
and toroidal, or {\it (2)} TeV-scale and toroidal, or {\it (3)} a
slice of AdS$_5$.

Of the above categories, only {\it (3)} allows for a curved
background.  This geometry is the basis of the Randall-Sundrum (RS) model
\cite{Randall:1999ee} which requires a 5-$d$ spacetime with constant
negative curvature.  Given that spheres provide a 
simple, highly symmetric, and yet non-trivial (with {\it positive}
constant curvature) departure from 
either the RS or toroidal geometries, 
it would be interesting to consider them as compactification 
manifolds.  TeV-scale phenomenology of
compactification on spheres  has so far received
very little attention; for some work in this direction see
Refs.~\cite{Lim,Leblond:2001ex}.  Perhaps this is due to the relative
simplicity of the analysis with a toroidal geometry, in conjunction
with the expectation that spherical extra dimensions would offer no
new qualitative features and only yield trivial numerical
modulations.

In this work, we consider models with extra dimensions compactified
on spheres $S^n$, $n \geq 2$, and show that the above expectation is
rather naive, since a number of new features will be shown to arise.
First of all, it has been demonstrated that if fermions propagate on
$S^n$, the low energy 4-$d$ spectrum does not include a chiral zero mode
\cite{Camporesi:1995fb,Abrikosov:2001nj}. For example, in the case
of $S^2$, the lightest Kaluza-Klein (KK) mode of a 6-$d$ fermion with a zero bulk
mass term has a mass $1/(2R)$, where $R$ is the radius of the sphere.{\footnote {In 
this case it can be shown that one can obtain a zero
mode KK fermion by a fine-tuning of the fermion bulk mass term. We
consider this possibility unnatural and will ignore it in the
discussion that follows.}} Hence, we are naturally led to a scenario
with ``non-universal" extra dimensions in which the fermions remain
4-$d$ fields. The easiest way to accomplish this is to have fermions
localized at some point on the surface of the $\sn$. {\it A priori},
all such points are identical by spherical symmetry and the arbitrary choice
of co-ordinates. Once a particular point is chosen for fermion
localization we can, for convenience of calculations, identify it
with one of the poles of the sphere. This fermion localization can
occur in a number of ways, \ie, through orbifolding and/or the
existence of a 3-brane at a particular point in the higher
dimensional space. The details of this particular mechanism are beyond 
the scope of the present analysis and will not concern us here.

In this paper, we will therefore examine the consequences 
of placing the gauge, and possibly Higgs,
sectors of the SM on a sphere.  Hence, we assume that $R^{-1} \gsim$~TeV,
to avoid conflict{\cite {LEPEWWG}} with low energy data.
\footnote{The KK graviton phenomenology with $R^{-1} \ll$~TeV and
an empty ``bulk" has been studied elsewhere \cite{Leblond:2001ex}.}  TeV-scale extra dimensions may play
a role in electroweak symmetry or supersymmetry breaking \cite{Antoniadis:1990ew}, and hence
may be discovered in upcoming LHC experiments \cite{colliders}.
We focus on the simple case with $n = 2$ where all the important features can be
inferred from the properties of the
familiar spherical harmonics.  Our results can be easily extended to the case of $n > 2$.  Upon
dimensional reduction, various
KK towers appear in the 4-$d$ description.  Due to the spherical
symmetry of the underlying geometry, there are preserved 
quantum numbers that ensure the stability
of some KK modes. As we will see, the $S^2$ spherical $SU(2)$ 
symmetry will be partially broken down to a
$U(1)$ by the localization of the fermions at the pole(s).

Exactly which KK states remain stable is a question of kinematics as well.  Here, quantum
corrections to the simple KK spectrum can affect our conclusions.  Also, we note that the stable
gauge sector modes
can become destabilized if the KK graviton spectrum allows ``gravitational" decays of the
gauge KK modes.  The resulting lifetimes are found to be reasonably long and these particles are
stable on collider time scales, but generally not cosmologically.  These issues have
also been noted in the previously studied toroidal compactifications \cite{Feng:2003nr} in the case of
models of Universal Extra Dimensions (UED){\cite{ued}}.  
We also briefly discuss the effects of radiatively generated
terms at the poles, induced by the localized fermions, in
the action for gauge fields.

The presence of pole-localized fermions in our setup provides SM production
and decay channels for certain KK resonances.  We use precision electroweak data
to constrain the mass of the lightest such KK modes to be larger than about 4~TeV.
Hence, it may be difficult to observe states beyond the first KK level even at LHC energies.  However,
we show that the features of a single KK resonance from spherical compactification are
quite distinct from those obtained from toroidal extra dimensions.  In addition, if the
second KK resonance is accessible, the mass ratio can be used to identify the underlying
spherical geometry.

Given that the geometries we study have $n \geq 2$ extra dimensions, there will be
one or more KK towers of physical scalars, corresponding to the the polarizations
of the gauge field along the sphere.  These scalars can be identified following standard
reduction procedures \cite{Carlson:2001bk}.  However, due to the non-trivial geometry,
this task is rather complicated in the spherical background.  Our study mainly focuses
on the phenomenology of the vector modes and their interactions with localized 4-$d$ fermion fields.  The
the spectrum and interactions of the scalars are omitted from our analysis since 
they do not play an important 
role in the discussion we present.  Nonetheless, we
expect that the KK scalars associated with the gauge sector further
enrich the phenomenology of spherical compactifications.

Our work is organized as follows. We will introduce the formalism related to spherical
compactification and our adopted
setup, in the next section.  Section III includes our discussion of the spectroscopy of these models.  Here, we
focus on the case with $n = 2$ and discuss the stability of the KK states.  We present the signatures of spherical
compactification at the LHC in section IV.  Our concluding remarks and a summary are included in the final section.

\section{Setup and Formalism}

In this section, we present our assumed physical setup and the relevant formalism that
we will employ in obtaining at our results.  We will consider a spacetime with $D=4 + n+m$, $n \geq 2$,
dimensions.  The $n$ extra dimensions are compactified on a sphere $S^n$ of radius $R$.
The other $m$ extra dimensions, if they exist, are assumed to be compactified on scales, $r$, which are
in general distinct from $R$.

Before considering placing the SM gauge fields on $\sn$, 
let us briefly examine possible effects from the gravity sector.    
The relation between the 4-$d$ (reduced) Planck scale \mpl ~and the 
fundamental (4 + $n$ + $m$)-$d$ scale $M_F$ is given by
\beq
\mpl^2 = M_F^{2 + n+m}\, V_{n+m},
\label{Mp2}
\eeq
where $V_{n+m} \sim R^nr^m$ is the product of the volumes of $\sn$ and all the other possible compact dimensions.  
As mentioned before, with bulk SM fields, we are 
led to assume $R^{-1} \gsim$~TeV, for consistency with low energy data{\cite {LEPEWWG}}.
{\it If} the above $m$ dimensions are assumed to have a size $M_F^{-1}$, then for $R^{-1}\sim$~TeV, we have
\beq
10^7~{\rm GeV} \lsim M_F \lsim 10^{11}~{\rm GeV},
\label{MF}
\eeq
with $2\leq n\leq6$.  Therefore, the gravity sector will not yield collider signatures in the TeV regime.
However, we will later discuss how KK gravitons affect the stability of the gauge KK modes.
Alternatively, with $M_F \gsim R^{-1}\sim$~TeV and $r \gg R$, we can reproduce many of the features
of the Arkani-Hamed, Dimopoulos, and Dvali (ADD) hierarchy model \cite{Arkani-Hamed:1998rs} and its
associated phenomenology.

Let us now consider the propagation of SM fields on $\sn$.
The first thing we note is that unlike with flat TeV-scale extra dimensions,
$\sn$ does not naturally accommodate SM fermions.  This is because the spectrum of
the Dirac operator on $\sn$ does not include a chiral zero mode and its lightest state has a mass
of order $R^{-1}$ \cite{Camporesi:1995fb}.  This problem cannot be resolved by 
orbifolding. We are, therefore, naturally led to exclude
fermions from propagating on the sphere.  For the rest of this work,
we will assume, as discussed above, that the fermion content of the
SM is localized at the pole(s) of $\sn$.  Hence, we will only consider the consequences of placing
the gauge/Higgs sectors of the SM on $\sn$ in what follows.  In this sense, the spherical extra dimensions
we are considering  are ``non-universal".

In studying the effects of spherical compactification in the gauge sector, we will mainly
focus on the case $n=2$.  This case encodes all the important features for any $n \geq 2$.
At the same time, the formalism and notation are simpler and more familiar for $n = 2$,
allowing for a more transparent presentation.  

The 6-$d$ metric for our setup is given by
\beq
ds^2 = \eta_{\mu\nu} \,dx^\mu dx^\nu - R^2\left[d\theta^2 + \sin^2\theta \,d\vphi^2\right],
\label{metric}
\eeq
where $0\leq \theta \leq \pi$ and $0\leq \vphi \leq 2 \pi$.  Here, we take the above 
geometry, with a flat 3-brane and extra dimensions compactified on a sphere, as a given.  
A proper derivation of this geometry, in accordance with Einstein's equations, requires 
the introduction of a bulk cosmological constant and a trapped Abelian gauge field, 
as demonstrated in Ref.~\cite{Leblond:2001ex}.  Once this background is obatined, we 
will treat the SM fields as  
weak perturbations that will not modify the underlying geometry, as is 
oft-assumed.  This is then the starting point of our analysis.

The action for a $U(1)$ gauge field in this spacetime
is given by
\beq
S = - \frac{1}{4}\int \!\!d^4x \!\int_0^\pi \!\!\!\!\! d\theta \!
\int_0^{2 \pi} \!\!\!\!\!\!\!d\vphi\, \sqrt{- g} \,\,g^{M R} g^{NS} F_{MN}F_{RS},
\label{S}
\eeq
where $\sqrt{- g} = R^2 \sin \theta$ and $M,N = 0, 1, \ldots, 5$.  For simplicity of notation, we
will henceforth mostly suppress powers of $R$ in
our presentation and only restore them for select final results.
One can expand the above action to get
\begin{eqnarray}\nonumber
S = \!\!\!\! & &- \frac{1}{4}\int \!\!d^4x \!\int_0^\pi \!\!\!\!\! d\theta \!
\int_0^{2 \pi} \!\!\!\!\!\!\!d\vphi\, \sin\theta
\{F_{\mu \nu}F^{\mu \nu} - 2[(\partial_\mu A_\theta - \partial_\theta A_\mu)
(\partial^\mu A_\theta - \partial_\theta A^\mu)\\
& &+ \sin^{-2}\theta
(\partial_\mu A_\vphi - \partial_\vphi A_\mu) (\partial^\mu A_\vphi - \partial_\vphi A^\mu)
- \sin^{-2}\theta (\partial_\theta A_\vphi - \partial_\vphi A_\theta)^2]\}.
\label{SII}
\end{eqnarray}
The second and third terms in the above expansion suggest that a linear combination of the fields $A_\theta$ and
$A_\vphi$ acts as a Goldstone boson to endow the $A_\mu$ KK tower with masses.
The orthogonal combination is left as a physical tower of scalars in the 4-$d$ effective theory.
This is familiar from the analysis of toroidal compactification.  However, a derivation leading
to the separation of the Goldstone and physical scalar towers is not as straightforward for $\sii$.
As we will focus on the gauge  vector boson KK phenomenology in our analysis, we do not consider these scalars
further in the following. {\footnote {The primary reason for doing this is that it can be
shown that such fields do not have zero modes \cite{Lim} nor do they
interact with the pole-localized SM fermions, as we will see below.}}
However, a more comprehensive treatment may include these KK scalars and their interactions with the vector KK
excitations of the SM gauge and Higgs fields.

Setting $A_\theta = A_\vphi = 0$ in Eq.~(\ref{SII}), and after integration by parts, we get
\beq
S = - \frac{1}{4}\int \!\!d^4x \!\int_0^\pi \!\!\!\!\! d\theta \!
\int_0^{2 \pi} \!\!\!\!\!\!\!d\vphi\,\sin\theta \left[-A_\mu \dalam A^\mu + \sin^{-1}\theta\, A_\mu\partial_\theta
(\sin\theta\partial_\theta A^\mu) + \sin^{-2}\theta \,A_\mu\partial^2_\vphi A^\mu\right],
\label{SAmu}
\eeq
where we have assumed the 4-$d$ gauge condition $\partial_\mu A^\mu = 0$ and defined $\dalam \equiv
\partial_\mu \partial^\mu$.  Solving the equation of motion corresponding to this action
yields the following solution for the KK expansion of $A_\mu$
\beq
A_\mu(x, \theta, \vphi) = \sum_{l=0}^\infty\sum_{m=-l}^l A_\mu^{(m,l)}(x) \,
\frac{Y^m_l(\theta, \vphi)}{R},
\label{AKK}
\eeq
where $Y^m_l(\theta, \vphi)$ are the familiar spherical harmonics on $\sii$.
As expected the $2 l + 1$ states with $-l \leq m \leq l$ for fixed $l$ are degenerate with a mass
\beq
m_l^2 = \frac{l(l+1)}{R^2}.
\label{ml2}
\eeq
These results are easily generalized to the non-Abelian case. 

We now consider the question of gauge sector KK interactions with the matter content of the SM.
The SM fermions are assumed to be localized at the poles on $\sii$, as discussed above.
The coupling of the generic fermions $\psi_1$ and $\psi_2$, localized at $\theta = 0, \pi$, respectively,
to the 6-$d$ gauge field $A_\mu$ is given by
\beq
S_f = \frac{g_6}{2}\int \!\!d^4x \!\int_{-1}^1 \!\!\!\!\! d(\cos\theta) \!
\int_0^{2 \pi} \!\!\!\!\!\!\!d\vphi
\left[{\bar \psi_1} A\!\!\!/(\theta, \vphi) \psi_1 \,\delta(\cos\theta - 1) +
{\bar \psi_2} A\!\!\!/(\theta, \vphi) \psi_2 \,\delta(\cos\theta + 1)\right],
\label{SfI}
\eeq
where $g_6$ is the gauge coupling constant which has mass dimension $-1$ and the factor of 1/2
accounts for the one-sided $\delta$-functions.  Since $A_\mu$ is expanded in terms of
$Y^m_l \propto e^{i m \vphi}$, we immediately see that only ``non-magnetic" states with $m=0$ have
non-zero coupling to the fermions $\psi_{1,2}$ which are localized
at $\theta=0,\pi$.  Using the expansion in (\ref{AKK}) and the explicit formula for $Y^0_l$, we find
\beq
S_f = \frac{g_6/2}{\sqrt{4 \pi R^2}}\int \!\!d^4x \left[{\bar \psi_1}\left(\sum_{l=0}^\infty
\sqrt{2 l + 1} \,A\!\!\!/^{(0,l)}(x) \right)\psi_1 + 
{\bar \psi_2}\left(\sum_{k=0}^\infty (-1)^k 
\sqrt{2 k + 1} \,A\!\!\!/^{(0,k)}(x) \right)\psi_2\right].
\label{SfII}
\eeq
We thus conclude that the 4-$d$ coupling $g_4$ of the zero mode $A^{(0,0)}_\mu$ to $\psi_{1,2}$ is given by
\beq
g_4 = \frac{g_6}{2\sqrt{4 \pi R^2}}.
\label{g4}
\eeq
The $A^{(0,0)}_\mu$ mode is to be identified as the corresponding conventional SM gauge field.
Thus the interaction in (\ref{SfII}) shows that the coupling of higher KK modes ($l > 0$)
to the localized fermions get progressively stronger
\beq
g^l_4/g_4 = \sqrt{2l + 1}.
\label{gl}
\eeq

As alluded to before, the fields $A_{\theta,\phi}$ do not have zero modes (\ie, have vanishing wavefunctions)
and will not couple to fermions. The reasons for this are easily seen: when $m\neq 0$ the KK wavefunctions 
for these fields behave as $\sim e^{im\phi}$
which clearly will not couple to pole localized states due to orthogonality. When $m=0$, the KK wavefunctions
for these fields are found to go as  $\sim \sin \theta$ \cite{Lim} which vanishes at both poles.

The localization of the fermions can lead to small shifts in the masses of these gauge boson KK states
through, \eg, the appearance of Pole-Localized Kinetic Terms (PLKT's){\cite {branes}}:
\beq
S_{PLKT} = \frac{1}{2} \int \!\!d^4x \!\int_{-1}^1 \!\!\!\!\! d(\cos\theta) \!
\int_0^{2 \pi} \!\!\!\!\!\!\!d\vphi \left[\alpha^i_0 F^i_{\mu \nu}F^{\mu\nu}_i \delta(\cos\theta - 1) +
\alpha^i_\pi F^i_{\mu \nu}F^{\mu\nu}_i \delta(\cos\theta + 1)\right],
\label{LKT}
\eeq
where $\alpha^i_{0,\pi}$ are gauge-group(labeled by the index `$i$') dependent constants in the
effective theory.  However, if the PLKT's are loop-generated, as
we will assume below, we then expect $\alpha^i_{0,\pi} \sim c_ig_{4i}^2log(R\Lambda)/(16 \pi^2)$ where the
$c_i$ are gauge-group dependent ${\cal O}(1)$ factors 
which explicitly depend on the localized fermion charges
and $\Lambda$ is some cutoff scale introduced to regulate the
loop-induced log divergence.  Since the localized fermions
only interact with the $m=0$ modes, the equation of motion for the $A_\mu$ KK wavefunction for these states $f_l$ is
now generically given by
\beq
\sin^{-1}\theta\,\partial_\theta(\sin\theta\,\partial_\theta f_l) + \left[1 +
\frac{\alpha_0}{2} \, \delta(\cos\theta - 1) +
\frac{\alpha_\pi}{2} \, \delta(\cos\theta + 1)\right] m_l^2 f_l = 0.
\label{EoMI}
\eeq
Away from the poles, $f_l = Y^0_l$ and we have
\beq
-l(l+1)\,Y^0_l + \left[1 +
\frac{\alpha_0}{2} \, \delta(\cos\theta - 1) +
\frac{\alpha_\pi}{2} \, \delta(\cos\theta + 1)\right]\,m_l^2 \,Y^0_l = 0.
\label{EoMII}
\eeq
Multiplying Eq.~(\ref{EoMII}) through by $Y^0_l$ and integrating over $\sii$, we find
\beq
m_l^2 = \frac{l(l+1)}{R^2 \left[1 + (\alpha_0 + \alpha_\pi) (2 l + 1)/4\right]}.
\label{ml2mod}
\eeq
For $\alpha_{0\pi} \ll 1$, the above ``perturbed" mass spectrum for the $m=0$ states is then well-approximated by
\beq
m_l^2 = \frac{l(l+1)}{R^2}\left[1 - (\alpha_0 + \alpha_\pi) (2 l + 1)/4\right],
\label{ml2pert}
\eeq
up to higher order ${\cal O}(\alpha_{0,\pi}^2)$ corrections, at low $l$.

The PLKT's pick out special points along the $\theta$-direction,
breaking the symmetry that protects $l$ conservation which results in mixing among states with $m = 0$.
Thus, the $Y^0_l$ are no longer the appropriate wave functions for the $m=0$ mass eigenstates,
as suggested by the mass formula (\ref{ml2pert}).  The new ``perturbed" eigenstates $\chi_l$ are then
directly given by Schr\"odinger perturbation theory to leading order:
\beq
\chi_l = Y^0_l - \sqrt{2l+1}\sum_{k\neq l}\frac{\sqrt{2k+1} \;Y^0_k}
{l(l+1) - k(k+1)} ~\left[\frac{\alpha_0 + (-1)^{k+l} \alpha_\pi}{4}\right] + {\cal O}(\alpha_{0\pi}^2),
\label{chil}
\eeq
with $l=0,1,2,\ldots$.  Here the index $l$ on $\chi_l$ is no longer a conserved quantum number
and merely enumerates the new eigenstates in mass order. It is important to note for later discussions
that all of the states $\chi_l$ now contain, \eg, a small component of $Y^0_0$.

The non-zero mode gauge KK fields also receive a common  but gauge-group dependent shift in their masses
from the finite size of the bulk as in the case of UED which can be expressed as
\begin{equation}
\delta m_i^2 \sim a_i {{g_{4i}^2}\over {16 \pi^2 R^2}}\,,
\end{equation}
where the numerical coefficients $a_i$ are essentially given by the Casimir invariants
of the relevant SM gauge group \cite{uedloops}.
This mass shift, together with the brane terms discussed above, make the gluon KK excitations heavier than
those of the electroweak gauge fields and the weak isospin fields heavier than the hypercharge fields as in UED.

Finally, we consider the Higgs sector.  With the SM fermions
localized at $\theta = 0, \pi$, it is most natural
to assume that the Higgs $H$ is also a 6-$d$ field.  The free action for $H$ can be written as \cite{Lim}
\beq
S_{H^2} = \int \!\!d^4x \!\int_0^\pi \!\!\!\!\! d\theta \!
\int_0^{2 \pi} \!\!\!\!\!\!\!d\vphi\,\sin\theta \left[-H^\dagger (\dalam - m_H^2) H +
\sin^{-1}\theta\, H^\dagger\partial_\theta
(\sin\theta\partial_\theta H) + \sin^{-2}\theta \,H^\dagger\partial^2_\vphi H\right],
\label{SHfree}
\eeq
with $m_H$ the mass parameter of the Higgs sector.  This action leads to a KK expansion
\beq
H(x, \theta, \vphi) = \sum_{l=0}^\infty\sum_{m=-l}^l H^{(m,l)}(x) \,
\frac{Y^m_l(\theta, \vphi)}{R}.
\label{HKK}
\eeq
for $H$ \cite{Lim}.  The 6-$d$ quartic term
$\lambda_6 (H^\dagger H)^2$, will then yield a 4-$d$ quartic
term for the zero mode $H^{(0,0)}$, identified as the SM Higgs,
with the coupling constant $\lambda = \lambda_6/(4 \pi R^2)$.
The mass term in (\ref{SHfree}) will yield a mass term $m_H^2 H^{(0,0)\dagger} H^{(0,0)}$.
The zero mode will then condense as usual in 4-$d$ and give
masses to the gauge field zero modes and fermions via Spontaneous Symmetry Breaking (SSB).

How does SSB via the Higgs vev modify the gauge KK masses? Clearly, level-by-level,
for the $SU(2)_L$ and $U(1)_Y$ gauge KK fields this SSB correction term induces a mass matrix
whose off-diagonal elements are of relative order $\sim (M_WR)^2$. If $R^{-1}\sim$ a few TeV, as will be
seen below, this SSB-induced mixing, \ie, the effective weak mixing angle for these states, can be
safely neglected to better than 1 part in 1000 on almost all occasions. Thus the KK
excitations of the $SU(2)_L$ and $U(1)_Y$ gauge fields can be treated as essentially unmixed, \ie, pure isospin or
hypercharge gauge KK excitations to an excellent first approximation,
which we will denote as $W^{0,\pm}$ and $B$, respectively.

As for the Yukawa couplings of the Higgs to fermions,
let us for simplicity consider the localized coupling to a fermion $\psi$ at $\theta = 0$,
\beq
S_{Y_6} = y_6 \int \!\!d^4x \!\int_{-1}^1 \!\!\!\!\! d(\cos\theta) \!
\int_0^{2 \pi} \!\!\!\!\!\!\!d\vphi
~H{\bar \psi_L}  \psi_R \,\delta(\cos\theta - 1),
\label{6dYukawa}
\eeq
where $y_6$ is the 6-$d$ Yukawa coupling.  Using the expansion (\ref{HKK}),
we then obtain the 4-$d$ interactions of the Higgs KK tower with $\psi$
\beq
S_Y = y \int \!\!d^4x ~{\bar \psi_L}\left(\sum_{l=0}^\infty
\sqrt{2 l + 1} \, H^{(0,l)}(x) \right)\psi_R,
\label{KKYukawa}
\eeq
where the 4-$d$ Yukawa coupling is $y = y_6/\sqrt{4 \pi R^2}$, identified as the SM
value for zero mode interactions. The absence of the fermions in the
bulk does not allow us to address the issue of the fermion mass hierarchy by localization.

\section{Spectroscopy and Lifetimes}

Given the mass spectrum and couplings of the SM gauge KK fields discussed above it is important to next
examine the `spectroscopy' of these various states. Much of this analysis can be obtained by rather
straightforward semi-quantitative considerations. As we will see, although there are some similarities to
the case of UED on $S^1/Z_2$ or $T^2/Z_2$, there are some important and very interesting differences. 
For the moment we will ignore gravitational interactions when discussing the
lifetimes and decay modes of the various gauge KK states.

As a prelude to this discussion we need to get a handle on the overall KK mass scale, ie, what is $R^{-1}$
or, in other words, what are the allowed masses of the lightest KK excitations. Bounds on the electroweak
gauge KK masses can be obtained by considerations of their effects on precision electroweak measurements{\cite {LEPEWWG}} 
as well as by constraints on possible contact interactions{\cite {precision}}. In the case of bulk Higgs fields,  
as is the case here, these
effects arise solely due to the additional KK exchanges which contribute to conventional SM amplitudes. These
contributions can be summarized in a single dimensionless parameter{\cite {precision}}
\begin{equation}
V=M_W^2\sum_k {{g_k^2/g^2}\over {M_k^2}}\,,
\end{equation}
where the, in principle infinite, sum extends over all KK states, labeled by the index $k$, coupling
to the localized SM fermions. Here, $g$ is just the zero-mode weak gauge coupling present in the SM. In the
well-studied case of $S^1/Z_2$, $g_k^2/g^2=2$ and $M_k=kM_1$ so that this sum converges yielding
$V=(\pi^2/3)(M_W^2/M_1^2)$. Thus, bounds on $V$ translate directly to bounds on $M_1$. Knowing
the bound in this case we can obtain the corresponding result for any other model through a simple rescaling.
In the
case of $S^2$, although only $m=0$ states contribute, the infinite sum is log divergent due to the growth
of the KK couplings with increasing $l$ found above. Of course, in practice we should only perform a sum over
a finite number of states as at some point the theory becomes strongly coupled. Due to the log behavior of
the sum, the resulting lower bound we obtain on the mass of the lightest KK state, 4-5 TeV, is only weakly
dependent on the employed cutoff. A similar situation is also seen to occur in the case of $T^2/Z_2$. 
Since these lightest KK states are so massive, it is clear that the effects
of SSB in the electroweak gauge can be generally neglected in discussing their associated physics.

When examining the lifetimes of the KK states, the most important feature to remember is that in all cases these decays 
will be prompt, \ie, all decays will occur essentially at the interaction point of the collider. The typical widths 
one finds are in the range of $\sim 0.01-1$ GeV so that if decays are allowed they occur rather quickly. Thus as far 
as signatures are concerned the actual lifetime values are not of immediate interest to us here. 
In order to understand the spectroscopy and lifetimes of the various KK states we can for the moment neglect
(almost) all of the correction terms to the zeroth order relationship given by Eq.~\ref{ml2} above except for
the effect of mass ordering within each KK level induced by loop effects. In this simple limit,
we can still label the various states by the integers $(l,m)$ recalling that the states with $m=0$ are only
approximate eigenstates of $l$. Let us first consider the lightest KK electroweak excitations which have
masses $M\simeq \sqrt {2}/R$. The states $(B,W^{0,\pm})_{l0}$~{\footnote {The raised indices here refer to the electric 
charge of the isotriplet gauge field; the lower indices refer to the $l$ and $m$ value of the particular state}}, 
since they have $m=0$, couple directly to the
localized SM fermions. In addition, they can be singly produced via collisions of and decay back into SM fermions in a rather
conventional manner. The state $B_{1\pm 1}$ is the lightest one with $m \neq 0$ and thus must be stable in the
limit where gravitational interactions are ignored due to the remaining $U(1)$ symmetry
and can be the LKKP as in UED models. The state $W^\pm_{1\pm 1}$ can then undergo a `2-body' decay as
$W^\pm_{1\pm 1}\to W^\pm_{SM}B_{1\pm 1}$; here, the SM $W$ field may be real or virtual depending upon the actual
numerical value of the mass splitting. Correspondingly, the neutral isotriplet state suffers a `3-body' decay as
$W^0_{1\pm 1}\to W^\mp_{SM}W^{\pm *}_{1\pm 1}+{\rm h.c.}\to W^+_{SM}W^-_{SM}B_{1\pm 1}$, with the $W$'s
again possibly being virtual. In the QCD sector, the gluon 
state $g_{10}$ couples to the SM localized quark fields at
the poles but the state $g_{1\pm 1}$ is now {\it stable}, unlike in UED, since it is neutral and
cannot couple to the localized quarks. Cosmologically, the stability of such a strongly interacting state
can be problematic{\cite {{Feng:2003nr,cosmo}}.

Let us now turn to the states with $l=2$. The first thing to notice is that (neglecting loop corrections) the
mass ratio of the $l=2$ to $l=1$ KK states is $\sqrt 3 <2$ so that on-shell decays of $l=2$ states into two
$l=1$ states is kinematically forbidden. Loop corrections are relatively small and do not change this result. To see the
overall pattern of decays for the $l=2$ level it is sufficient to consider the case of the gluon KK states; the
patterns for the $W^{0,\pm}$ gauge fields can be analyzed in an analogous fashion employing the discussion in the
previous paragraph whereas the $B$ states will present a special case we return to below. The $g_{20}$ KK couples to
fermions and can be produced and decay in the usual manner. $g_{2\pm 1}$ does not couple to fermions but can decay
via the gauge non-Abelian trilinear coupling:  $g_{2\pm 1}\to g_{1\pm 1}g_{l0}^*$, with $l=1,2,$ or 3 in the limit of 
exact $l$ conservation, with  $g_{l0}^*$ representing
either the the virtual state with these ($l,m$) values or the zero mode gluon field which can appear via mixing.
In either case the $g_{l0}^*$ can decay to pole localized fermions. Similar arguments will apply to the
$W^{0,\pm}_{2\pm 1}$ KK fields since they also have trilinear couplings. In the $B$ case,  $B_{20}$ can decay
by direct fermion couplings whereas $B_{2\pm 1}$ requires a trilinear coupling to decay; such a coupling is
absent if this state is a {\it pure} hypercharge excitation. Fortunately, SSB induces a tiny mixing with
$W^0_{2\pm 1}$ via an effective Weinberg angle of order $\sim 10^{-3}-10^{-4}$. This mixing generates a small trilinear 
coupling so that the state $B_{2\pm 1}$ can decay. 

The states $W^{0,\pm}_{2\pm 2}$ have a more serious problem as the only 
potential decay path  is via stable modes, \eg, $W^0_{2\pm 2}\to W^\mp_{1\pm 1}W^\pm_{1\pm 1}$ which is forbidden 
by kinematics. In principle, the $W^\pm_{1\pm 1}$ states can go off shell, however, their decay chain 
ends in a stable $B_{1\pm 1}$ state. Since the mass difference between the $W_1$ and $B_1$ states is 
loop-generated, and hence small, we see that this decay cannot proceed via intermediate off-shell states.  
Thus the KK modes $W^{0,\pm}_{2\pm 2}$ must be stable. By an identical argument one sees that $g_{2\pm 2}$ are 
{\it also} stable. Furthermore, in a similar vein one can easily demonstrate that all of the states 
$(g,B,W^{0,\pm})_{l\pm l}$ are stable, which could be cosmologically problematic \cite{cosmo}, 
whereas all other heavy KK states can decay directly to fermions or via 
trilinear couplings that may be SSB mixing induced.

As suggested in Ref.~\cite{Feng:2003nr} such potential cosmological problems can be 
circumvented by recalling that we have ignored gravitational
interactions. Though the gravitons, $G$, may exist in more extra dimensions than on the sphere, those
along the sphere 
will couple to all of the SM gauge fields allowing for their decay. For any set of ($l,m$) these gravitons are
the quite likely to be the 
lightest states and the lightest one of which will now play the role of the LKKP. Since the relevant
couplings are Planck scale, these lifetimes can be quite long, differing from all the decays discussed above in an 
important qualitative way. As an example, consider the typical decay of
this kind $B_{1\pm 1}\to G_{1\pm 1}\gamma$. The width for this decay can be calculated to be\cite{Feng:2003nr}
\begin{equation}
\Gamma={{\cos ^2 \theta_W M_B^3}\over {72\pi \mpl^2}}\big[ x^{-2}(1-x)^3(1+3x+6x^2)\big]\,,
\end{equation}
where $\mpl$ is the 4-$d$ reduced Planck scale and $x=M_G^2/M_B^2$ 
in obvious notation. Defining the mass difference $\Delta=M_B-M_G$, we can calculate the
rest frame $B_{1\pm 1}$ lifetime as shown in Fig.~\ref{fig1}. For typical splittings such lifetimes can be
measured in days or weeks. Thus  $B_{1\pm 1}$ will be stable on collider scales but not on cosmological scales.
\begin{figure}[htbp]
\centerline{
\includegraphics[width=7.5cm,angle=90]{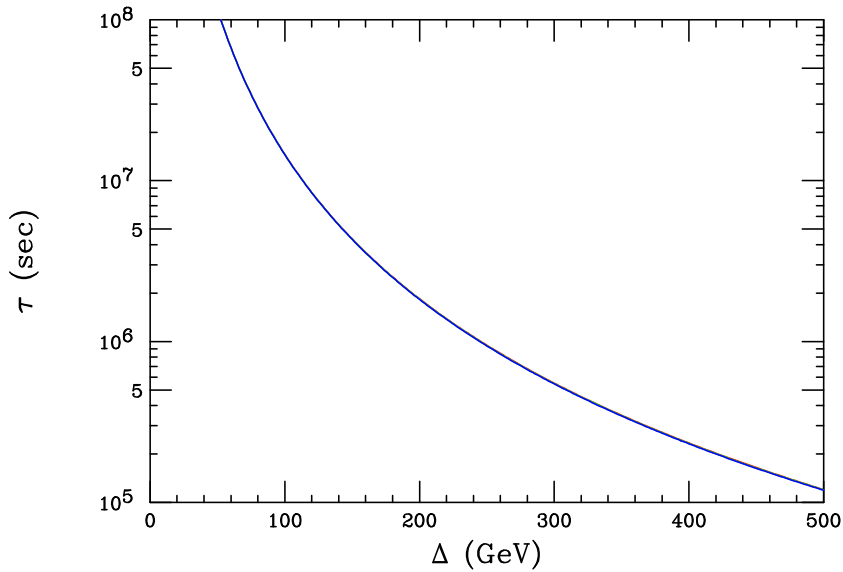}}
\vspace*{0.1cm}
\caption{Lifetime for the decay $B_{1\pm 1}\to G_{1\pm 1}\gamma$ as a function of the mass splitting, for 
$M_G=3-5$~TeV.  Note that the $M_G$ mass dependence within the thickness of the curve cannot be resolved over the plot's range.}
\label{fig1}
\end{figure}
Other gravitationally induced lifetimes can be obtained in a similar fashion with qualitatively similar results.

\section{Collider Signatures}

The next questions to address are ($i$) can we see the physics associated with this $S^2$ picture and ($ii$) can
we differentiate the present model from, \eg, the more conventional scenarios such as $S^1/Z_2$ and $T^2/Z_2$? To answer 
them we must be able to directly observe the KK excitations of the
various SM gauge bosons at the LHC. Except for possible kinematic limitations, 
the resonant $m=0$ KK states should be accessible
in a straightforward manner. The observation of pair produced KK states seems to be 
very unlikely due to their large masses which greatly 
suppress their production cross sections{\cite {pairs}} even at LHC energies.

The resonant neutral electroweak KK states can be produced in the Drell-Yan channel $q\bar q \to
X \to l^+l^-$; here $X$ includes the SM zero modes $\gamma$ and $Z$ as well as all of the kinematically accessible
KK states \cite{colliders}. In the discussion that follows we will assume that all of the SM fermions are located 
at {\it either} the North or South Poles of the sphere, \ie, have either $\theta=0$ or $90^o$ but not both. 

\begin{figure}[htbp]
\centerline{
\includegraphics[width=7.5cm,angle=90]{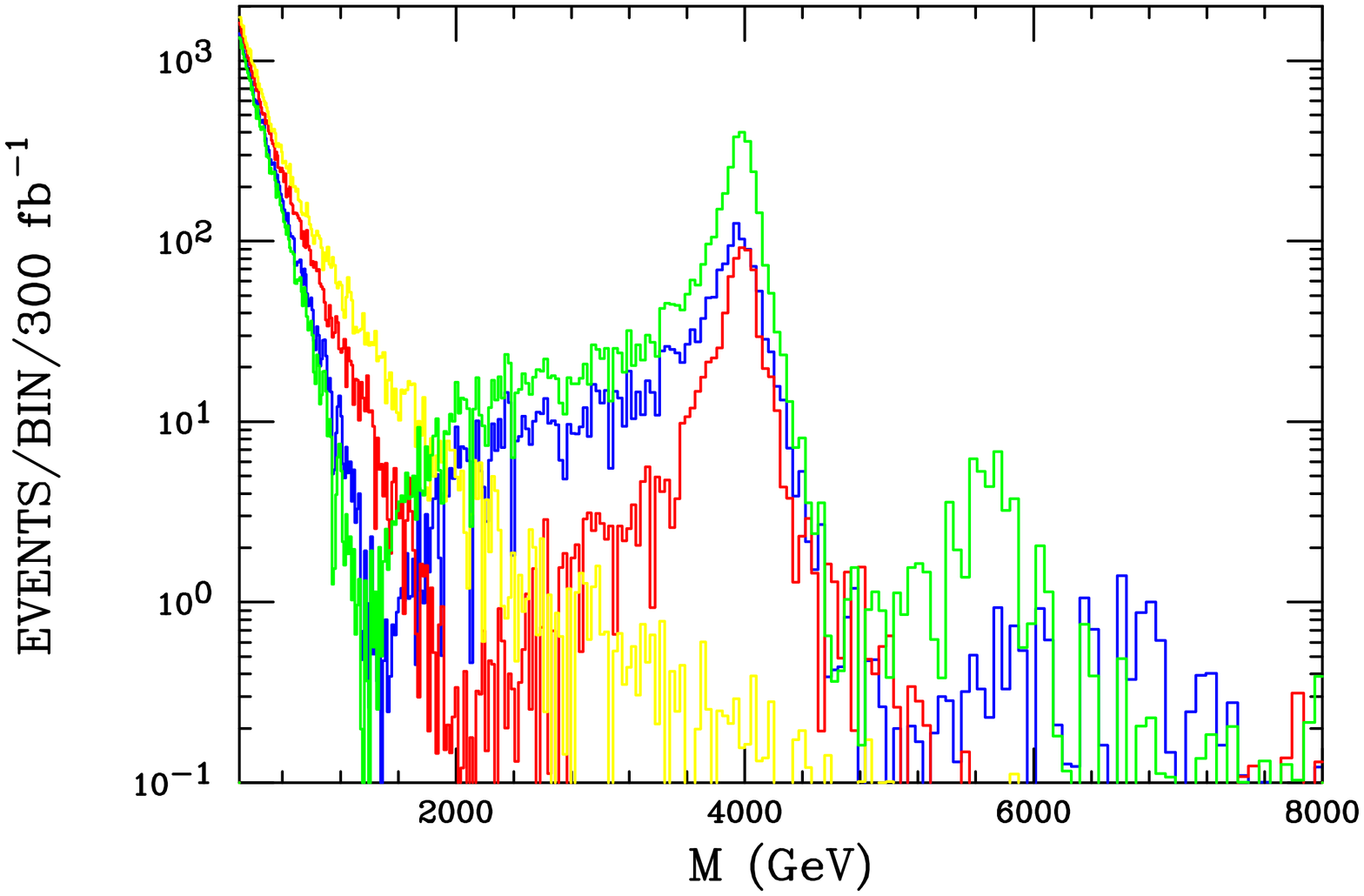}}
\vspace*{0.1cm}
\centerline{
\includegraphics[width=7.5cm,angle=90]{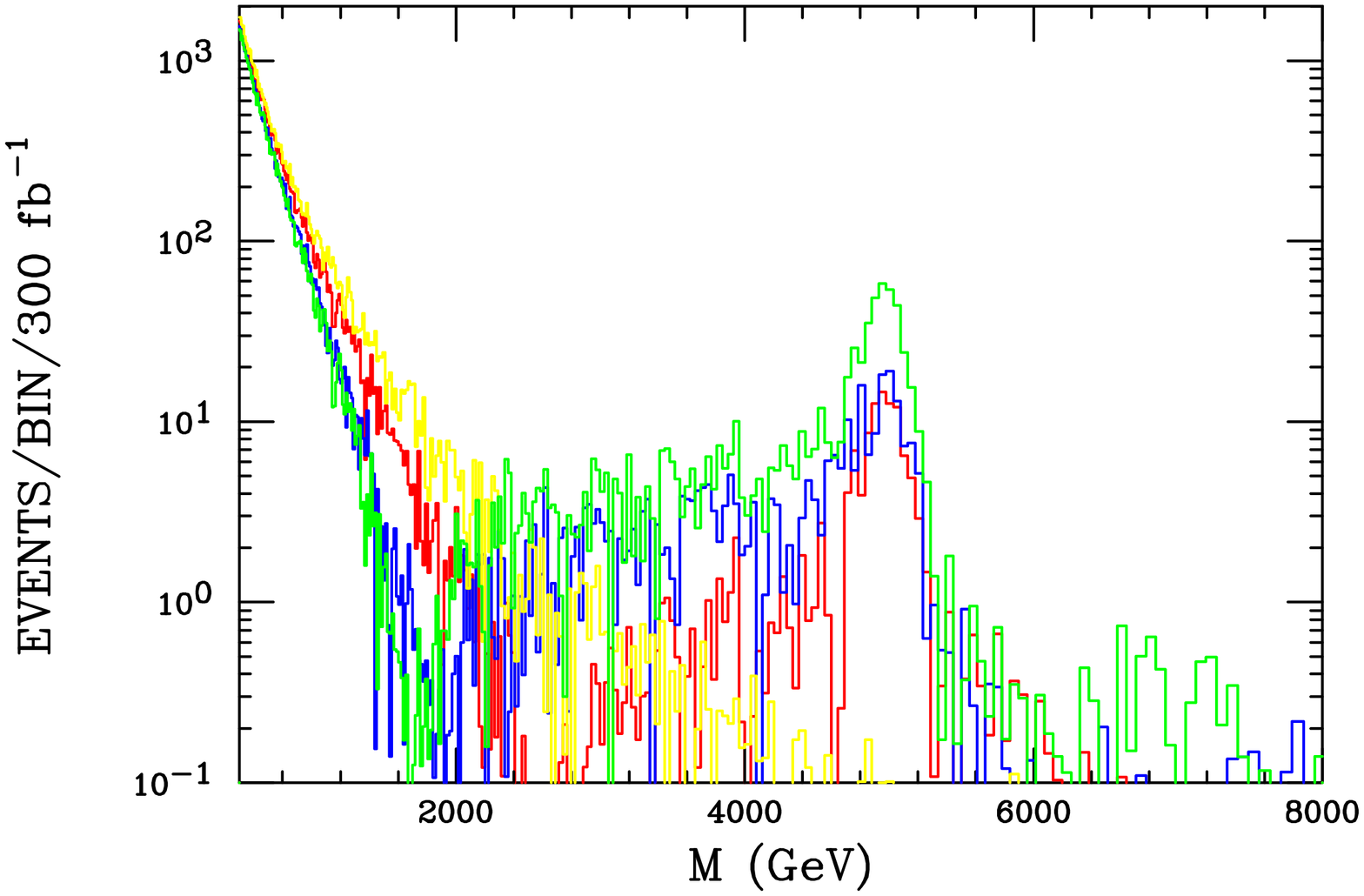}}
\vspace*{0.1cm}
\caption{Drell-Yan production rate as a function of the dilepton pair invariant mass of the neutral electroweak
KK resonances at the LHC. The upper(lower) plot corresponds to
the case where the lightest KK has a mass of 4(5) TeV. The yellow histogram in both panels is the SM background while
the blue(red,green) histogram corresponds to the case of $S^2(S^1/Z_2,T^2/Z_2)$. Fermions are 
assumed to be completely localized at either the North or South 
Poles. The bin size is $1\%$ of the dilepton invariant mass.}
\label{fig2}
\end{figure}

Fig.~\ref{fig2} shows the case where the lightest KK mass is assumed to be 4 or 5 TeV. For either mass value
the degenerate KK resonance structure due to the simultaneous production of the states $(B,W^0)_{10}$ can be
observed above the SM backgrounds. At first glance looking at the Figure, this may not appear to be the case. However, we 
must remember that these are binned distributions. For example, in the case of $M=4$ TeV, if we make a cut of 2 TeV 
on the minimum dilepton pair mass we find 96 SM induced background events and 1670(763,4007) signal events in the case of the 
$S^2(S^1/Z_2,T^2/Z_2)$ model. In the case of $M=5$ TeV, there would be 1454(309,1717) signal events for these two models. 
There is thus no question of the presence of a signal in all cases. 

Comparing, $S^2$ with $S^1/Z_2$, due to the larger couplings, the resonance structure of the $S^2$ case is
significantly broader and the well-known \cite{colliders} KK destructive interference minimum occurs at a 
significantly lower value of the dilepton invariant mass. $T^2/Z_2$ is also distinctive due to both the double degeneracy of the 
first KK level, which produces a generally larger cross section, and the relatively low mass of the second KK excitation.  
These differences are all clearly visible in the 4 TeV case but are somewhat
harder to observe in the case where the first KK state has a mass of 5 TeV assuming an integrated luminosity of 300 $fb^{-1}$. 
Unlike in the $T^2/Z_2$ scenario, in either mass case for $S^2$ or $S^1/Z_2$ it is unlikely that
any higher resonances due to more massive KK states would be observed. This situation improves significantly if we
consider the luminosity upgrade of the LHC \cite{Gianotti:2002xx} as shown in Fig.~\ref{fig3}. {\footnote {If for no 
other reason, such an
upgrade would naturally occur if the KK resonance structures discussed here were to be observed.}}
Here we see that if the first $S^2$ KK mass lies at 4 TeV, it may be possible with higher luminosity to observe the
degenerate $(B,W^0)_{20}$ structure at a mass of $\simeq 6.93$ TeV, which is predicted to be more massive, \ie, 8
TeV in the case of the $S^1/Z_2$ scenario. For the $T^2/Z_2$ scenario, the second KK state lies at $\simeq 5.66$ TeV and 
is easily visible even at the lower luminosity. 
This further aids in distinguishing the three models. It is clear that at higher 
luminosities, the lightest KK states may be observable up to $\simeq 7$ TeV or more for all classes of models.  
\begin{figure}[htbp]
\centerline{
\includegraphics[width=7.5cm,angle=90]{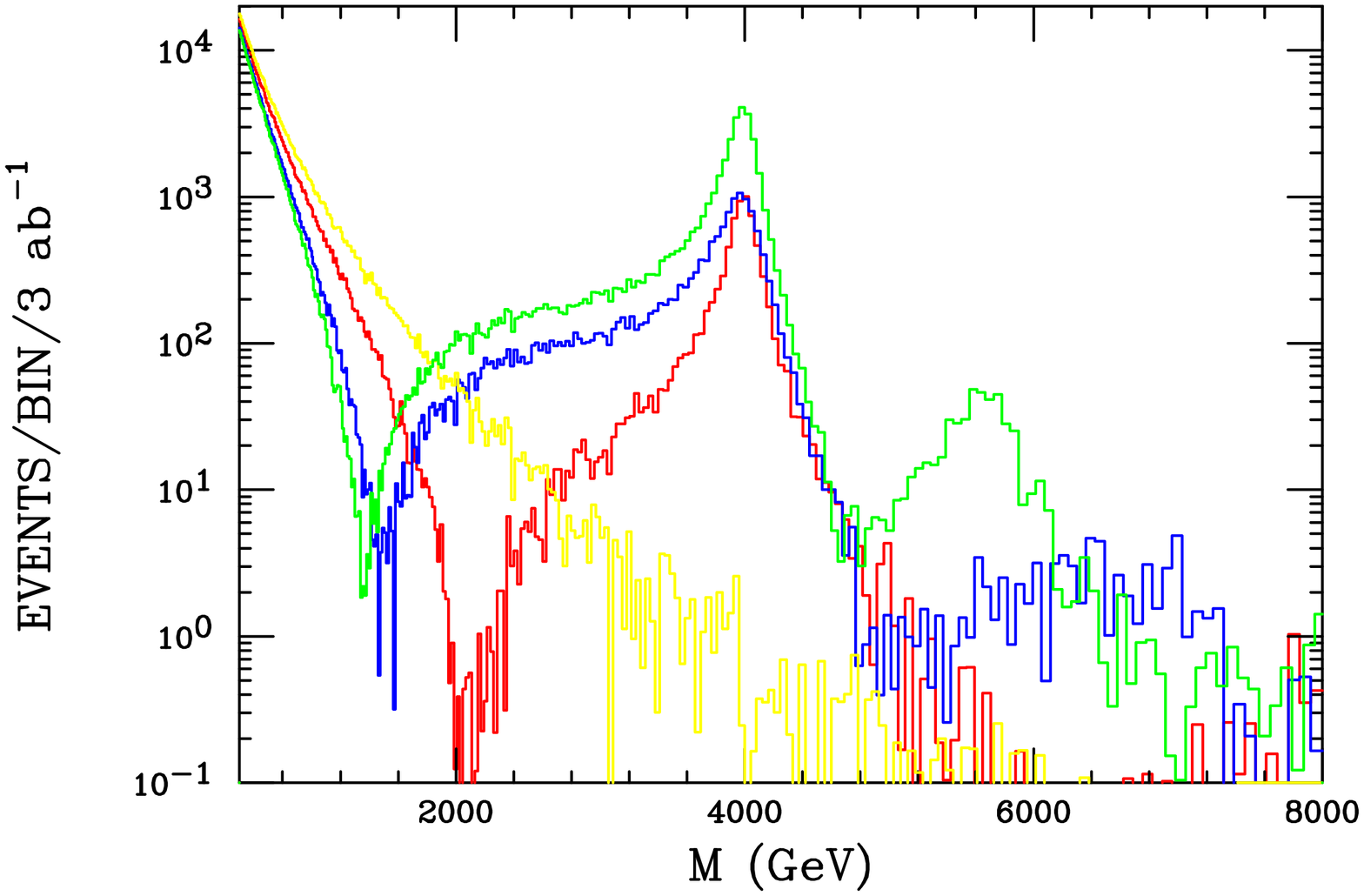}}
\vspace*{0.1cm}
\centerline{
\includegraphics[width=7.5cm,angle=90]{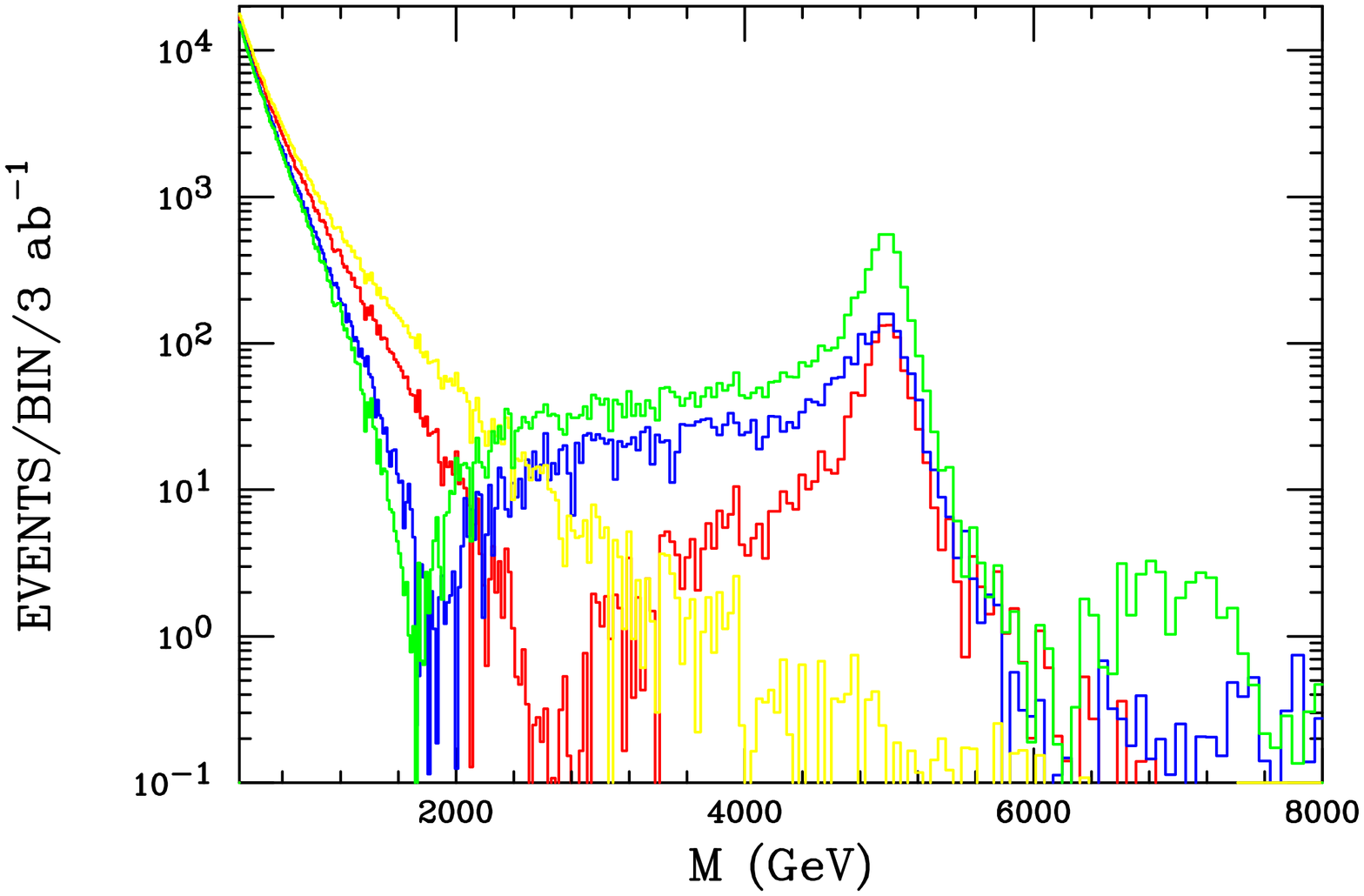}}
\vspace*{0.1cm}
\caption{Same as the previous figure but now for the LHC upgrade with an 
order of magnitude higher integrated
luminosity.}
\label{fig3}
\end{figure}

In order to help verify that any new resonances observed at 
the LHC are due to the production of KK excitations of the SM gauge fields, 
it is necessary to see their production in several channels. 
In addition to the $q\bar q \to (B,W^0)_{l0}\to l^+l^-$ channel discussed 
above, the lightest of the corresponding $W^\pm_{l0}$ states 
should also be observed via the process $q\bar q' \to W^\pm_{l0}\to l^\pm +E_T^{miss}$ \cite{colliders}. 
Fig.~\ref{fig3p} shows the transverse mass distribution 
for the lepton plus missing $E_T$ final state induced by 
the production of these charged states at the LHC. It is clear from 
this Figure that the direct production of these states should 
most likely be 
visible out to $\simeq 6$ TeV in this channel and that $S^2-S^1/Z_2$ 
model differentiation is possible for masses up to approximately 5 TeV. Differentiation of $S^2$ and 
$T^2/Z_2$ is seen to be significantly more difficult in this channel even with the high integrated luminosities 
assumed here. 
\begin{figure}[htbp]
\centerline{
\includegraphics[width=7.5cm,angle=90]{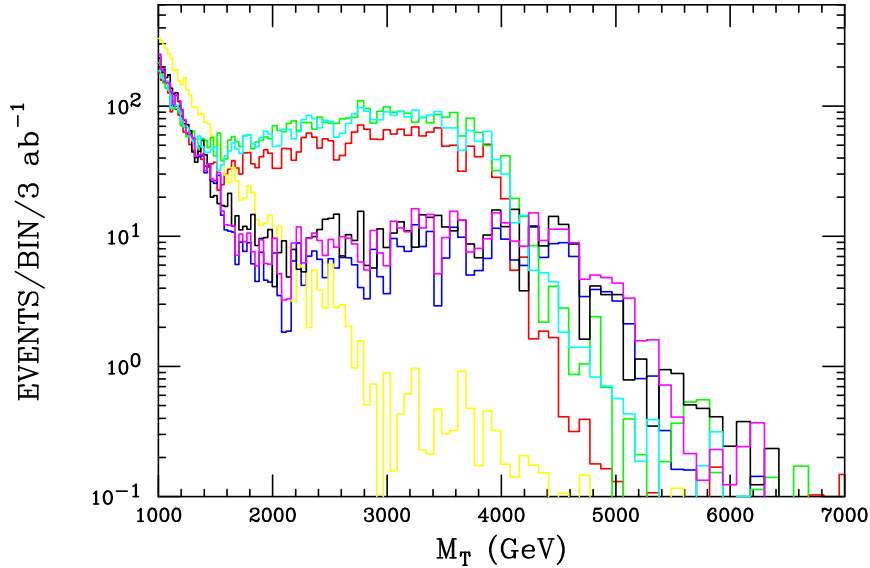}}
\vspace*{0.1cm}
\caption{Transverse mass distributions for $W^\pm_{l0}$ production at the LHC. The lowest, steeply falling histogram is the SM 
background. The top(lower) triplet of signal histograms is for a lightest $W^\pm_{10}$ KK mass of 4(5) TeV. The 
upper(middle,lower) member of each pair is for the $T^2/Z_2$($S^2$,$S^1/Z_2$)model.}
\label{fig3p}
\end{figure}

KK signatures can also be observed in other channels associated with the gluon KK excitations $g_{l0}$. 
Fig.~\ref{fig4} shows the dijet mass spectrum at the LHC subject to
a pair of selection cuts for centrally produced objects. 
Here, we see that the existence of gluon KK excitations does not produce a
significant resonance structure, except in the $T^2/Z_2$ case where there is added constructive interference from the 
degenerate pair of KK states. This is due to the fact that these resonances are 
rather wide and their relative cross sections, being in the $q\bar q$
channel, are relatively suppressed. {\footnote {Note that in the limit where any 
mixing of the $m=0$ KK states can be neglected there is no coupling
between the zero mode SM gluons and the various KK states. We make this assumption in the numerical 
results presented here.}} 
Instead, one generally sees a rather broad shoulder induced by the
existence of these KK states. For all of 
these models this shoulder should be observable 
for gluon KK masses in excess of 7 TeV. In the specific case
where the lightest gluon KK mass is 4 TeV, the height of 
this shoulder is seen to be significantly different for the $S^2$ and 
$S^1/Z_2$ models allowing them to be easily distinguished. It remains difficult in this 
channel to distinguish the $S^2$ and $T^2/Z_2$ models away from the peak region. 
However, we note that, \eg, a 5 TeV KK in the $S^1/Z_2$ model produces a signal which
is quite similar to a 7 TeV KK in the $S^2$ case. 
Thus, in general, since there is no obvious resonance structure for the $S^2$ and $S^1/Z_2$ models, 
they are only distinguishable in this channel if the mass 
of the lightest KK excitation is already known from other measurements such
as the Drell-Yan channel discussed above.
\begin{figure}[htbp]
\centerline{
\includegraphics[width=7.5cm,angle=90]{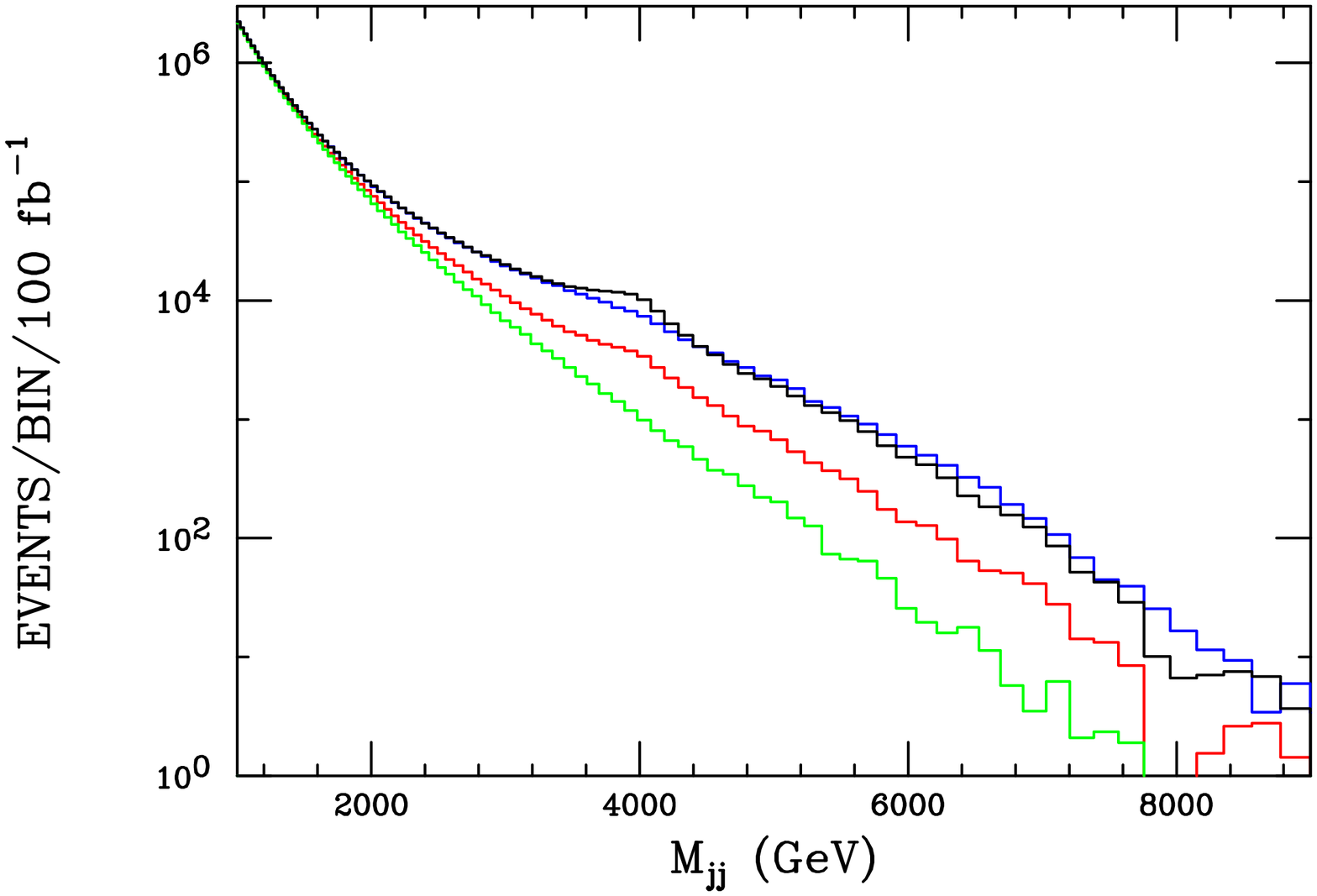}}
\vspace*{0.1cm}
\centerline{
\includegraphics[width=7.5cm,angle=90]{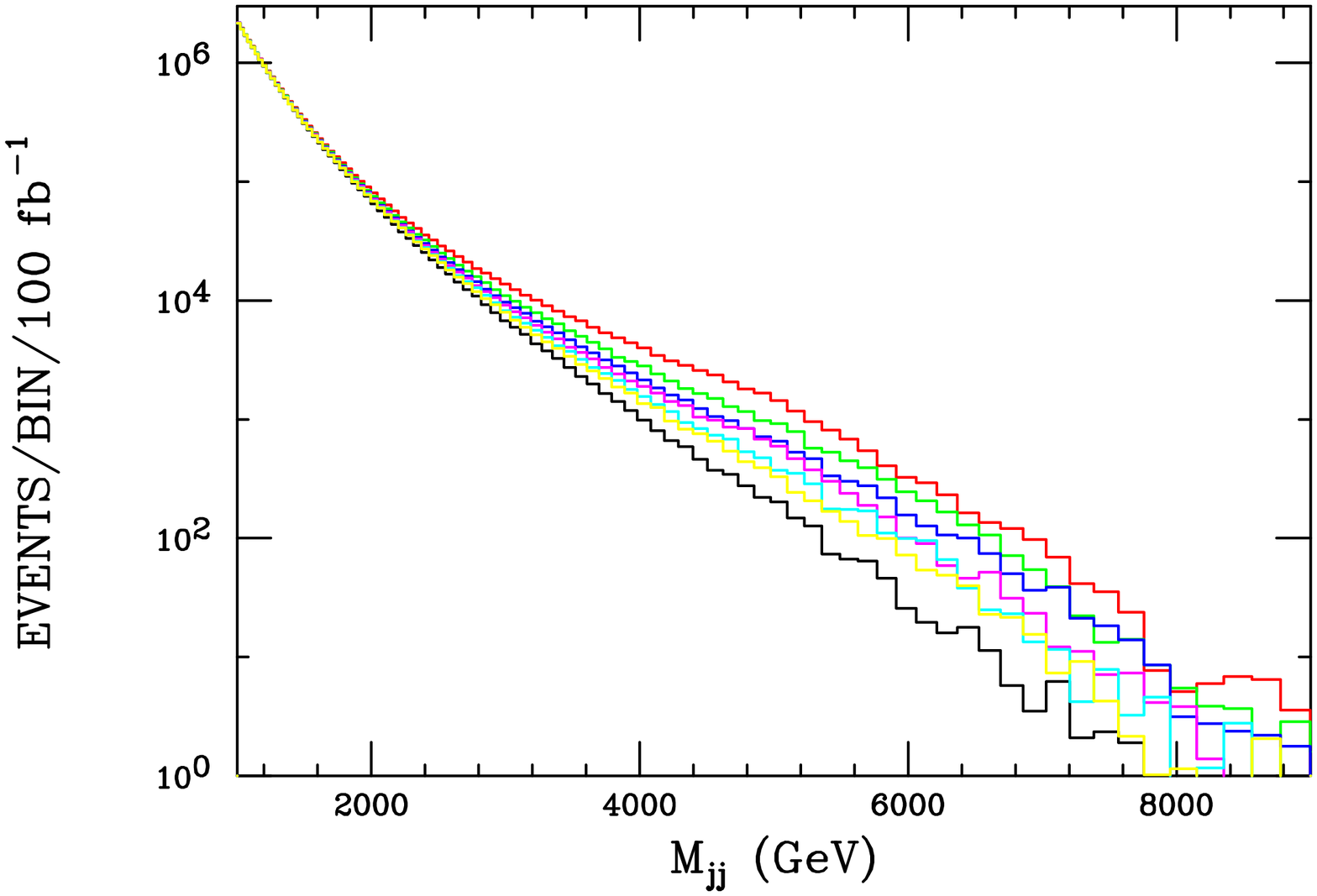}}
\vspace*{0.1cm}
\caption{Dijet pair mass at the LHC applying the cut $|\eta| \leq 1$ and $p_{Tj}\geq 0.4 M_{jj}$. The lowest histogram
in both panels is the SM QCD background. In the top panel, a $g_{10}$ mass of 4 TeV has been assumed with the
upper(middle,lower) signal histogram corresponding, on the right-hand side of the figure, 
to the $S^2(T^2/Z_2,S^1/Z_2)$ model. In the lower panel the top(middle) three histograms
are for the $S^2(S^1/Z_2)$ model assuming, from top to bottom, KK masses of 5, 6 or 7 TeV, respectively. Couplings between
the KK states and SM zero mode gluons have been neglected.}
\label{fig4}
\end{figure}

Mixing between the SM zero mode gluon and the corresponding $m=0$ gluon KK excitations, while inducing a $ggg_{l0}$ coupling,
does not alter these results in any significant way since the induced coupling is quite small being of order
$\sim g_{4s}^3 log^2 (\Lambda R)/128\pi^3$.  
Even though the $gg$ luminosity is generally larger than $q\bar q$ 
luminosity, it cannot compensate at such large $x$ 
values for the rather strongly suppressed loop-induced $ggg_{l0}$ coupling; this is especially true 
at larger dijet masses which are relevant here.

From the above discussion, it is clear that the KK states of 
interest will be visible in all resonant channels and that the $S^2$ model of interest to 
us here can be differentiated from both the $S^1/Z_2$ and $T^2/Z_2$ model cases  
provided that the mass scale for these states is not too large.

\section{Conclusions}

In this work, we studied a scenario in which the SM gauge 
and possibly Higgs sectors propagate in compact spherical 
extra dimensions.  Since spheres $S^n$ do not allow chiral zero 
modes for fermions, these particles are naturally 
assumed to be localized at the poles and remain 
4-dimensional in this scenario.  The fermions can then lead to the 
appearance of pole localized kinetic terms for the gauge sector, 
which can result in level mixing among the ``non-magnetic" ($m = 0$) 
KK modes.  

We focused on $S^2$ 
as a simple representative case and analyzed the 
vector KK towers of the gauge sector.  We found that the 
symmetries of the geometry result in the appearance 
of certain KK gauge fields that may be stable.  This 
picture can change once KK gravitons are assumed to 
be the be lightest states, level by level, as they are 
expected to receive suppressed quantum corrections to 
their masses.  In that case, the previously stable 
states can decay into KK gravitons with macroscopically 
long lifetimes.  Other gauge KK modes can decay into SM 
fermions which are localized at poles, and hence such states may be produced as  
resonances at the LHC.  

Current precision electroweak bounds push the mass of the the first 
KK mode to about 4~TeV.  However, we have shown that a 4-5~TeV KK mode 
is well within the reach of the LHC.  The features of these resonances 
are quite distinct from their toroidal ($S^1/Z_2$ and $T^2/Z_2$) counterparts.  
A luminosity-upgraded LHC with 3~ab$^{-1}$ delivered can potentially access the  
second KK excitations of the SM gauge fields and establish the characteristic 
mass ratios special to $S^2$.

In general, we find that the gauge KK resonances on $S^2$ can be distinguished from 
those of other compactifications, such as $S^1/Z_2$ and $T^2/Z_2$, 
by the ratios of KK masses, the growing strength of the couplings of the KK fields to 
pole localized SM fermions, and the size of the production cross section.  
In particular, the lightest KK mode on $S^2$ couples 
to SM fermions with enhanced strength compared to its $S^1/Z_2$ counterpart.  This feature 
affects the shape of the KK resonance in all the channels we examined.  The first 
$T^2/Z_2$ KK mode has the largest production cross section of all three cases, due to double 
degeneracy.  The second KK mode is lightest in the $T^2/Z_2$, 
and heaviest in the $S^1/Z_2$ case, given the same compactification radius.  

In summary, we showed that the properties of spherical extra dimensions are qualitatively 
different from the more familiar toroidal ones.  These new features can 
lead to the emergence of alternative models and open novel avenues for phenomenological studies.

\acknowledgments

H.D. is supported in part by the DOE grant
DE-AC02-98CH10886 (BNL).

\end{document}